# High-harmonic generation from an epsilon-near-zero material

Yuanmu Yang,[1*] Jian Lu[2], Alejandro Manjavacas[3], Ting S. Luk,[4,5] Hanzhe Liu[2], Kyle Kelley,[6] Jon-Paul Maria,[6] Michael B. Sinclair,[4] Shambhu Ghimire[2], Igal Brener[4,5*]

[1]State Key Laboratory for Precision Measurement Technology and Instruments, Department of Precision Instrument, Tsinghua University, Beijing 100084, China

[2]Stanford PULSE Institute, SLAC National Accelerator Laboratory, Menlo Park, California, 94025, USA

[3]Department of Physics and Astronomy, University of New Mexico, Albuquerque, New Mexico 87131, USA

[4]Sandia National Laboratories, Albuquerque, New Mexico 87185, USA

[5]Center for Integrated Nanotechnologies, Sandia National Laboratories, Albuquerque, New Mexico 87185, USA

[6]Department of Materials Science and Engineering, North Carolina State University, Raleigh, North Carolina 27695, USA

*Corresponding authors: ymyang@tsinghua.edu.cn or ibrener@sandia.gov

**High-harmonic generation (HHG) from a compact, solid-state medium is highly desirable for applications such as coherent attosecond pulse generation and extreme ultra-violet (EUV) spectroscopy[1], yet the typically weak conversion of pump light to HHG can largely hinder its applications. Here, we use a material operating in its epsilon-near-zero (ENZ) region[2-11], where the real part of its permittivity vanishes, to greatly boost the efficiency of the HHG process at the microscopic level. In experiments, we report high-harmonic emission up to the 9th order directly from a low-loss, solid-state ENZ medium: indium-doped cadmium**

**oxide, with an excitation intensity at the GW cm⁻² level. Furthermore, the observed HHG signal exhibits a pronounced spectral red-shift as well as linewidth broadening, resulting from the photo-induced electron heating and the consequent time-dependent resonant frequency of the ENZ film. Our results provide a novel nanophotonic platform for strong field physics, reveal new degrees of freedom for spectral and temporal control of HHG, and open up possibilities of compact solid-state attosecond light sources.**

While traditionally observed in rare-gas atoms[12], HHG has also recently been reported in solids, with reduced threshold pump field[13-16] and the additional advantage of producing stable EUV waveforms[17] in a compact setup. Unfortunately, above-band-gap absorption and short coherence lengths restricts the HHG process to a very thin layer of the solid-state material (typically tens of nanometers in thickness), significantly limiting the generation efficiency[18].

Nanostructures with plasmonic resonances are widely known to provide enhancements of the optical near field, and therefore have been investigated for HHG in their surrounding media[19,20]. However, plasmonic nanostructures are also known to exhibit low damage threshold under intense laser fields and the electric field enhancement is restricted to the near fields at the metal-dielectric interface. Furthermore, the periodic arrangement of plasmonic antennas leads to diffraction of the generated high-harmonic signal over a wide range of angles[16,20], making the generation and detection of high harmonics unwieldy. More recently, materials that exhibit a vanishing real part of their permittivity in certain spectral ranges, commonly known as ENZ materials, have been found to exhibit outstanding nonlinear optical properties[2-11]. Although not yet explored, ENZ materials are appealing for HHG since the ENZ effect can provide a large optical nonlinearity in a robust, un-structured, deep sub-wavelength film, thus alleviating the issue of above-band-gap absorption of the high harmonics. Moreover, upon ultrafast high field excitation, the unique hot electron dynamics of the ENZ material[2,10,21] may lead to intriguing spectral and temporal control of the high-harmonics.

In this work, we report harmonic generation up to the 9$^{th}$ order directly from an ENZ material: indium (In)-doped cadmium oxide (CdO). The structure, as illustrated in Fig. 1a, is a Berreman-type plasmonic cavity[22,23] formed from a 75-nm-thick In-doped CdO layer and 200-nm-thick gold capping layer, sequentially deposited on a magnesium oxide (MgO) substrate. The In-doped CdO film has a measured carrier density of $2.8 \times 10^{20}$ cm$^{-3}$ and an electron mobility of 300 cm$^2$V$^{-1}$s$^{-1}$. According to the Drude dispersion formula for this material, the real part of its permittivity crosses zero at a wavelength of 2.1 µm. The deep sub-wavelength thickness of the CdO film is comparable to its skin depth at the ultraviolet wavelengths of the high-harmonic radiation (Supplementary Information).

The origin of the field enhancement near the material's ENZ wavelength can be intuitively understood by the continuity of normal displacement field $\varepsilon_{MgO}E_{MgO} = \varepsilon_{CdO}E_{CdO}$ ($\varepsilon_{MgO}$ and $\varepsilon_{CdO}$ are the permittivity of MgO and CdO, respectively, and $E_{MgO}$ and $E_{CdO}$ are the normal electric fields at the material boundary). When the real part of $\varepsilon_{CdO}$ vanishes, $E_{CdO}$ diverges in the limit that the imaginary part of $\varepsilon_{CdO}$ approaches zero. Compared to conventional transparent conducting oxides such as indium-doped tin oxide, CdO benefits from the ~5 times lower imaginary permittivity at the ENZ wavelength due to its high electron mobility[24,25], leading to an unrivaled resonant field enhancement. Furthermore, the ENZ wavelength can be varied by adjusting the doping level of the CdO film[24]. Figure 1b shows the field enhancements as a function of the wavelength and incidence angle calculated from the transfer matrix method[26], showing an intensity enhancement of up to sixteen-fold when the pump wavelength is tuned to the ENZ wavelength. Moreover, as illustrated in Fig. 1c, the field enhancement drops off sharply in the regions outside the ENZ material, indicating that the nonlinear optical response of the structure should be boosted dominantly in the ENZ material itself.

We generate high harmonic radiation by illuminating the sample at an angle of incidence $\theta$ with linearly polarized laser pulses of 60 fs pulse duration and with a 1 kHz repetition rate. The emitted harmonic radiation propagates along the specular reflection direction, collinear with the pump beam (Methods). Figure 2a shows the high-harmonic spectra obtained from the layered structure under *p*- and *s*-polarized illumination at $\theta = 50°$, with the excitation wavelength $\lambda_0$ centered at 2.08 µm. For *p*-polarized illumination, odd-order harmonics are observed from the 3$^{rd}$ to the 9$^{th}$ order for an excitation peak intensity of 11.3 GW cm$^{-2}$. Due to the large resonant field enhancement in the ENZ material, this excitation intensity is much weaker than the typical intensity of ~1 TW cm$^{-2}$ used for HHG from zinc oxide and silicon[13,15]. The greatly reduced pump threshold could potentially allow the observation of HHG in the ENZ material from a femtosecond oscillator system. In contrast, for *s*-polarized illumination, a configuration that does not lead to field enhancements, no harmonic signal above the 3$^{rd}$ order is observed. The measured 3$^{rd}$ harmonic yield for *p*-polarized illumination is 482 times greater than that for *s*-polarized illumination. These results serve as clear evidence that HHG originates in the CdO, enhanced by the ENZ effect. In Fig. 2b, we show the harmonic yield as a function of the excitation intensity. While the 3$^{rd}$ harmonic for *s*-polarized illumination scales as $I^3$, where $I$ is the incident laser intensity (an indication of a perturbative effect), the 3$^{rd}$ to the 7$^{th}$ harmonics for *p*-polarized illumination all deviate from perturbative scaling and exhibit a gradual saturation. The harmonic yield is reversible after repeated measurements, indicating that the measurements do not damage the film (Supplementary Information).

To confirm the resonant nature of HHG in the ENZ sample, we measure the 5$^{th}$ harmonic yield with the central wavelength of the excitation pulse tuned across CdO's ENZ wavelength of 2.1 μm under an identical intensity of 5.6 GW cm$^{-2}$, as illustrated in Fig. 3a. Despite the as-expected resonant enhancement of harmonic yield near the ENZ wavelength, we note that the peak of the harmonic yield red-shifts to an incident wavelength of 2.13 μm.

Upon careful examination of the harmonic spectra in Fig. 2a, we find that all the harmonic peaks exhibit substantial, excitation intensity-dependent, red-shifts away from the expected harmonic photon energy $N\hbar\omega_0$, the product of the fundamental photon energy $\hbar\omega_0$ and the harmonic order $N$, where $\hbar$ is the reduced Planck constant. The measured $N^{th}$ harmonic peak appears approximately at $N\times 0.93\hbar\omega_0$ for all harmonic orders. In Fig. 3b, we present the measured 5$^{th}$ harmonic spectra from the ENZ sample as functions of the excitation intensity along with the 5$^{th}$ harmonic spectrum from a bare silicon wafer. The resonantly-enhanced harmonic peak from the ENZ material exhibits a pronounced red-shift as the excitation intensity increases, and the red-shift saturates at an excitation intensity of 11.3 GW cm$^{-2}$. Moreover, when comparing with the harmonic spectrum from a bare silicon wafer which is an indicator of the fundamental excitation bandwidth, we observe that the resonantly-enhanced harmonic peak exhibits a broadening as large as 1.83 times (shown in Fig. 3c). This unique feature of linewidth broadening could be used to tailor or reduce the temporal duration of HHG pulses. At moderately high excitation intensities, a second spectral peak corresponding to the non-resonant harmonic generation emerges at a photon energy close to $5\hbar\omega_0$.

We attribute the microscopic origin of the spectral red-shift and broadening of the harmonic radiation to the photo-induced electron heating and the consequent time-dependent resonant frequency of the CdO film, as schematically illustrated in Fig. 4a. Upon sub-bandgap optical excitation, conduction band electrons in CdO are heated to a much higher temperature $T_e$. The electron heating leads to a modulation of the effective electron mass in CdO and a red-shift of CdO's ENZ wavelength, according to the Drude formula[21,27]. The time scale of the modulation is sub-picosecond, and is comparable to the dwell time of the excitation pulse inside the CdO cavity[11], which leads to a unique feature in our system that the excitation pulse interacts with a time-variant cavity while emitting harmonics. This is manifested in the observed red-shift and linewidth broadening of the time-averaged HHG intensity profile. Here, we provide a simplified model to calculate the harmonic spectral peak and bandwidth at a given excitation intensity (Supplementary Information). In the model, the pump-modified ENZ wavelength of CdO is determined only by the maximum attainable $T_e$, while the actual amount of frequency shift and broadening shall be less pronounced, governed by the average $T_e$ within the time duration that the excitation pulse interacts with the CdO cavity. In Fig. 4b, we show the calculated 5$^{th}$ harmonic spectra with increasing excitation intensities, which shows progressively increasing spectral red-shift and broadening in qualitative agreement with the experimental results. We observe similar spectral red-shift and broadening in the 3$^{rd}$ harmonic spectra with *p*-polarized incidence (Supplementary Information).

To summarize, we have opened a new paradigm of utilizing the ENZ effect for HHG directly from a solid-state material and explored the unique spectral shift and broadening associated with the ultrafast hot electron dynamics of the ENZ material. Looking forward, a broadened high-harmonic signal could suggest a pathway towards reduced pulse duration approaching the attosecond regime. The temporal and spectral properties of the harmonic emission can be further manipulated by introducing a frequency chirp to the pump pulse and matching the instantaneous laser frequency to the temporal evolution of the ENZ frequency. HHG in CdO could also be coherently controlled with multiple beams simultaneously perturbing the structure and emitting harmonics. The ENZ frequency of CdO is tunable by adjusting its doping level either chemically or through electrostatic gating. Finally, conductive metal oxides including CdO with ENZ responses have found their ways to be integrated with silicon photonics technology[28], outscoring the possibility of compact, on-chip sources for EUV and attosecond pulses.

**Acknowledgements**

This work performed at Sandia National Laboratories was supported by the US Department of Energy (DOE), Office of Basic Energy Sciences, Division of Materials Sciences and Engineering, and performed, in part, at the Center for Integrated Nanotechnologies, an Office of Science User Facility operated for the U.S. Department of Energy (DOE) Office of Science. Sandia National Laboratories is a multi-mission laboratory managed and operated by National Technology and Engineering Solutions of Sandia, LLC., a wholly owned subsidiary of Honeywell International, Inc., for the U.S. DOE's National Nuclear Security Administration under contract DE-NA-0003525. The views expressed in the article do not necessarily represent the views of the U.S. DOE or the United States Government. The work performed at SLAC was primarily supported by the US Department of Energy, Office of Science, Basic Energy Sciences, Chemical Sciences, Geosciences, and Biosciences Division through the Early Career Research Program. Y.Y. acknowledge support from the Youth Thousand Talent program of China. A.M. acknowledges the National Science Foundation (Grant ECCS-1710697), as well as the UNM Center for Advanced Research Computing for providing the high performance computing resources used in this work. S.G. thanks Uwe Thumm for stimulating discussions.

**Author Contributions**

Y.Y. conceived the idea; J.L., Y.Y., T.S.L and H.L. carried out optical measurements; A.M. developed the theory; K.K. and J.P.M. prepared the sample; all authors contributed to analyzing the data and writing the manuscript; I.B., S.G., M.B.S. and J.P.M. supervised the project.

**Competing Financial Interests**

The authors declare no competing financial interests.

## Methods

**HHG measurements.** The experimental setup for measuring the HHG from the ENZ material is shown schematically in the supplementary information. The light source is an optical parametric amplifier (OPA, HE-TOPAS-Prime, Light Conversion) pumped by a Ti:sapphire amplifier system (Legend Elite Duo, Coherent) operating at a 1 kHz repetition rate. The 60-fs idler pulse from the OPA which is tunable from 2.0 to 2.2 μm in wavelength is loosely focused into the sample at a ~50 °angle of incidence. The area of focal spot on the CdO surface is estimated to be ~0.0297 $cm^2$ from a knife edge measurement. The high-harmonic beam propagating along the specular reflection direction is collected by a lens and focused into the entrance slit of a spectrometer, which consists of a monochromator (Acton, SP-300i) and a charge-coupled device (CCD, Andor). Spectrally-resolved high-harmonic signals emerging from the ENZ material are recorded by the CCD. The polarization of the idler pulse is tuned by a half wave-plate and the excitation intensity is controlled by a variable neutral density (ND) filter. The reported spectral signals are not corrected for the efficiency of the spectrometer at different wavelengths or for *s*- or *p*-polarized pump pulses.

## Data availability

The data that support the plots within this paper and other findings of this study are available from the corresponding author on reasonable request.

**Figures and Legends**

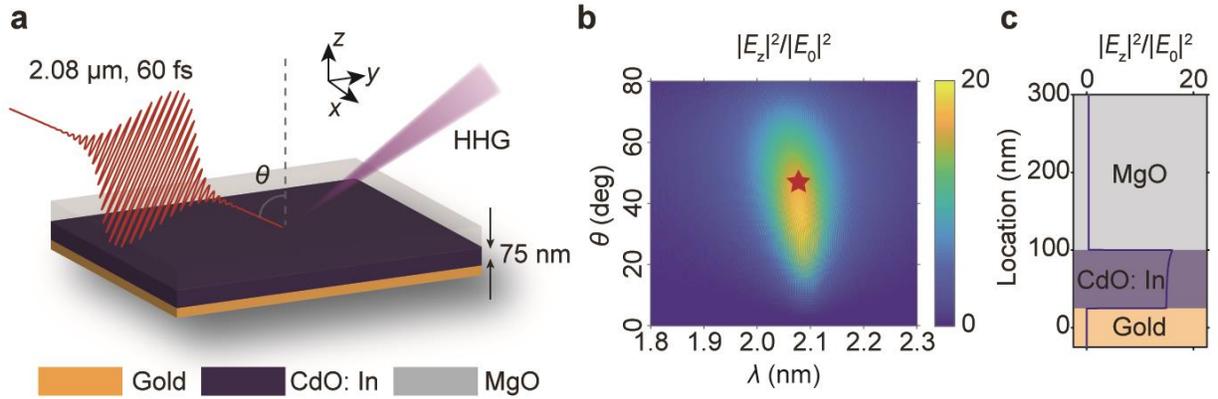

**Figure 1| Sample schematics and linear optical responses. a**, Schematics of the ENZ sample composed of a MgO substrate, CdO thin film and gold capping layer. The pump laser is incident from the substrate side with an incident angle of $\theta$ in free space. **b**, The field enhancements in CdO for *p*-polarized light as a function of wavelength and incident angle calculated from a transfer matrix model. $E_0$ is the incident electric field in free space. **c**, The calculated field enhancements of the three-layer structure as a function of location at a wavelength $\lambda = 2.08$ μm, and $\theta = 50°$ (marked with a red star in panel b).

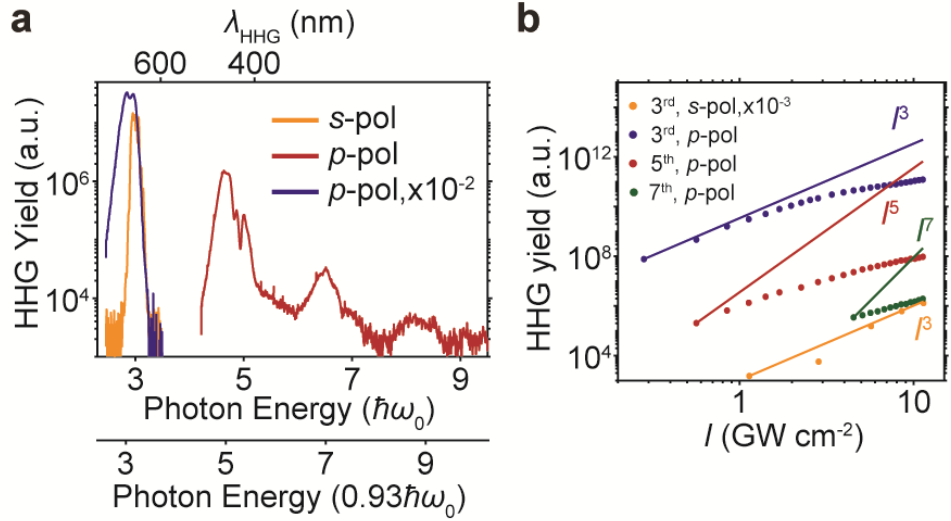

**Figure 2| HHG from CdO. a**, High-harmonic spectrum of the CdO-based structure. For *p*-polarized illumination, the spectrum extends from 3$^{rd}$ to the 9$^{th}$ order (blue and red line). For *s*-polarized illumination, only 3$^{rd}$ harmonic generation is observed (orange line). $\hbar\omega_0 = 0.6$ eV is the pump photon energy. **b**, Harmonic yield as a function of pump intensity. For *p*-polarized illumination, the 3$^{rd}$ to the 7$^{th}$ harmonics all scale non-perturbatively (blue, red, and green dot). For *s*-polarized illumination, the 3$^{rd}$ harmonic generation scales perturbatively (orange dot). The solid lines are plotted to guide the eye.

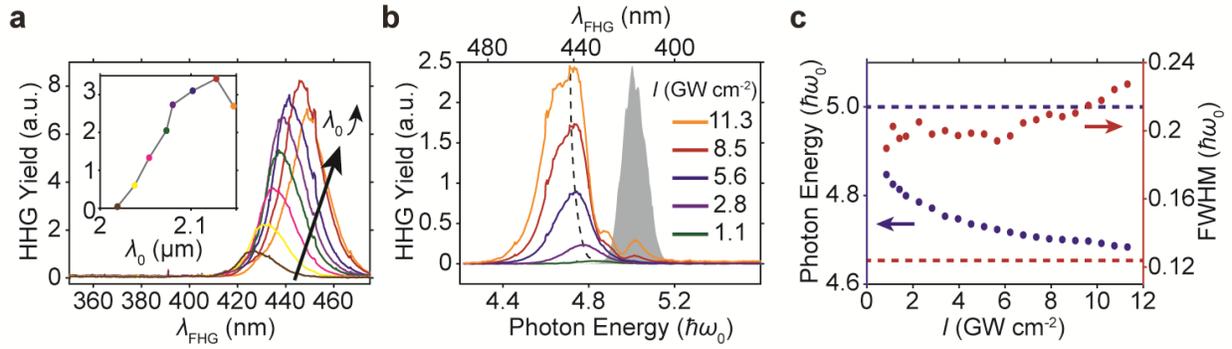

**Figure 3| Harmonic spectral shift and broadening. a**, 5th harmonic spectra as a function of the excitation wavelength $\lambda_0$, with an identical pump intensity of 5.6 GW cm$^{-2}$. The $\lambda_0$ of each spectrum is specified by the dot with corresponding color in the inset. The inset also shows the integrated harmonic yield as a function of $\lambda_0$. **b**, Measured spectra of 5th harmonic generation (FHG) for *p*-polarized illumination as a function of the pump intensity. The black dashed line is plotted to guide the eye, showing the red-shift of the harmonic peaks with increasing pump intensity. The gray shaded area marks the 5th harmonic from a silicon crystal subjected to a pump intensity of 54.2 GW cm$^{-2}$. **c**, Harmonic spectral peak location (blue dot) and spectral full width at half maximum (FWHM) (red dot) as a function of the pump intensity. The harmonic spectral peak location (blue dashed line) and spectral FWHM (red dashed line) from a silicon crystal are also plotted for comparison.

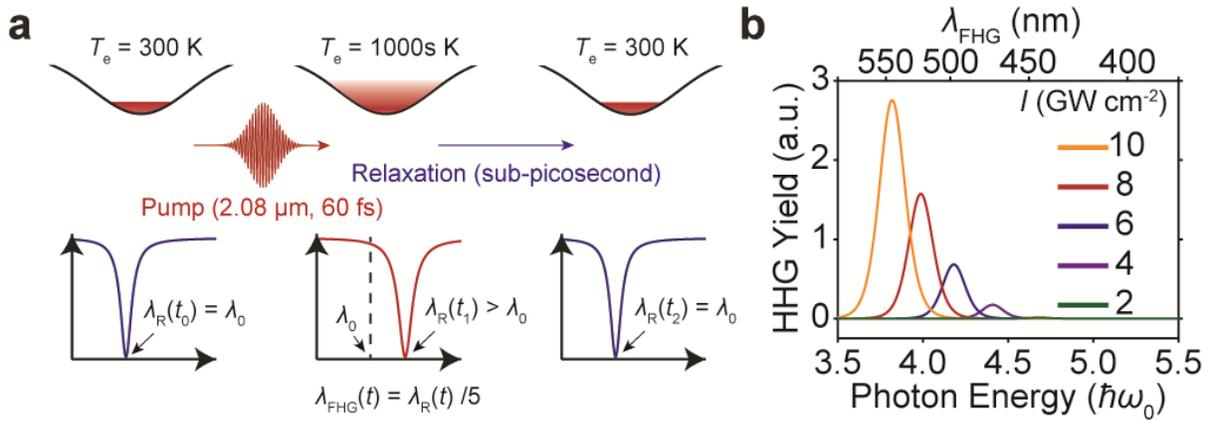

**Figure 4| Theoretical interpretation. a**, Schematics of electron heating and resonance wavelength shift upon photo-excitation. $\lambda_R$ denotes the ENZ resonance wavelength of the cavity, which is a function of time $t$. Here, $t_0 < t_1 < t_2$. **b**, Calculated spectra of FHG for *p*-polarized illumination as a function of the pump intensity.